\documentclass[aip, amsmath, amssymb,reprint]{revtex4-2}

\usepackage[utf8]{inputenc}
\usepackage{graphicx}
\usepackage{dcolumn}
\usepackage{bm}
\usepackage[utf8]{inputenc}
\usepackage[T1]{fontenc}
\usepackage{mathptmx}
\usepackage{gensymb}
\usepackage{upgreek}
\usepackage{xcolor}
\usepackage{textcomp}
\usepackage{hyperref}
\usepackage{mhchem}
\usepackage{float}
\usepackage[verbose]{placeins} 
\hypersetup{colorlinks=true,linkcolor=red,citecolor=blue,filecolor=magenta,urlcolor=magenta}

\usepackage{color,soul} 

\usepackage{titlesec}
\titleformat{\section}[hang]{\small\bfseries\sffamily}{\thesection.}{0.5em}{\MakeUppercase}
\titlespacing{\section}{0pc}{1pc}{0.2pc}

\begin{document}
\title{CMOS-compatible photonic integrated circuits on thin-film ScAlN}

\author{Sihao Wang}
\thanks{These two authors contributed equally}
\affiliation{Institute of Materials Research and Engineering (IMRE), Agency for Science, Technology and Research (A$^\ast$STAR), 2 Fusionopolis Way, Innovis \#08-03, Singapore 138634, Republic of Singapore}
\author{Veerendra Dhyani}
\thanks{These two authors contributed equally}
\affiliation{Institute of Materials Research and Engineering (IMRE), Agency for Science, Technology and Research (A$^\ast$STAR), 2 Fusionopolis Way, Innovis \#08-03, Singapore 138634, Republic of Singapore}
\author{Sakthi Sanjeev Mohanraj}
\affiliation{Institute of Materials Research and Engineering (IMRE), Agency for Science, Technology and Research (A$^\ast$STAR), 2 Fusionopolis Way, Innovis \#08-03, Singapore 138634, Republic of Singapore}
\author{Xiaodong Shi}
\affiliation{Institute of Materials Research and Engineering (IMRE), Agency for Science, Technology and Research (A$^\ast$STAR), 2 Fusionopolis Way, Innovis \#08-03, Singapore 138634, Republic of Singapore}
\author{Binni Varghese}
\affiliation{Institute of Microelectronics (IME), Agency for Science, Technology and Research (A$^\ast$STAR), \\2 Fusionopolis Way, Innovis \#08-02, Singapore 138634, Republic of Singapore}
\author{Wing Wai Chung}
\affiliation{Institute of Microelectronics (IME), Agency for Science, Technology and Research (A$^\ast$STAR), \\2 Fusionopolis Way, Innovis \#08-02, Singapore 138634, Republic of Singapore}
\author{Ding Huang}
\affiliation{Institute of Materials Research and Engineering (IMRE), Agency for Science, Technology and Research (A$^\ast$STAR), 2 Fusionopolis Way, Innovis \#08-03, Singapore 138634, Republic of Singapore}
\author{Zhi Shiuh Lim}
\affiliation{Institute of Materials Research and Engineering (IMRE), Agency for Science, Technology and Research (A$^\ast$STAR), 2 Fusionopolis Way, Innovis \#08-03, Singapore 138634, Republic of Singapore}
\author{Qibin Zeng}
\affiliation{Institute of Materials Research and Engineering (IMRE), Agency for Science, Technology and Research (A$^\ast$STAR), 2 Fusionopolis Way, Innovis \#08-03, Singapore 138634, Republic of Singapore}
\author{Huajun Liu}
\affiliation{Institute of Materials Research and Engineering (IMRE), Agency for Science, Technology and Research (A$^\ast$STAR), 2 Fusionopolis Way, Innovis \#08-03, Singapore 138634, Republic of Singapore}
\author{Xianshu Luo}
\affiliation{Institute of Microelectronics (IME), Agency for Science, Technology and Research (A$^\ast$STAR), \\2 Fusionopolis Way, Innovis \#08-02, Singapore 138634, Republic of Singapore}
\author{Victor Leong}
\affiliation{Institute of Materials Research and Engineering (IMRE), Agency for Science, Technology and Research (A$^\ast$STAR), 2 Fusionopolis Way, Innovis \#08-03, Singapore 138634, Republic of Singapore}
\author{Nanxi Li}
\email{linx1@ime.a-star.edu.sg}
\affiliation{Institute of Microelectronics (IME), Agency for Science, Technology and Research (A$^\ast$STAR), \\2 Fusionopolis Way, Innovis \#08-02, Singapore 138634, Republic of Singapore}
\author{Di Zhu}
\email{dizhu@nus.edu.sg}
\affiliation{Institute of Materials Research and Engineering (IMRE), Agency for Science, Technology and Research (A$^\ast$STAR), 2 Fusionopolis Way, Innovis \#08-03, Singapore 138634, Republic of Singapore}
\affiliation{\parbox{15cm}{%
Department of Materials Science and Engineering, National University of Singapore, Singapore 117575, Singapore
}}
\affiliation{\parbox{15cm}{%
Centre for Quantum Technologies, National University of Singapore, Singapore 117543, Singapore
}
}


\begin{abstract}
Scandium aluminum nitride (ScAlN) has recently emerged as an attractive material for integrated photonics due to its favorable nonlinear optical properties and compatibility with CMOS fabrication. 
Despite the promising and versatile material properties, it is still an outstanding challenge to realize low-loss photonic circuits on thin-film ScAlN-on-insulator wafers. 
Here, we present a systematic study on the material quality of sputtered thin-film Sc$_{0.1}$Al$_{0.9}$N produced in a CMOS-compatible 200 mm line, including its crystallinity, roughness, and second-order optical nonlinearity, and developed an optimized fabrication process to yield 400 nm thick, fully etched waveguides. 
With surface polishing and annealing, we achieve micro-ring resonators with an intrinsic quality factor as high as $1.47\times 10^5$, corresponding to a propagation loss of 2.4 dB/cm. 
These results serve as a critical step towards developing future large-scale, low-loss photonic integrated circuits based on ScAlN.

\end{abstract}
\maketitle

\section{Introduction}

Photonic integrated circuits (PICs) hold great promise in realizing compact and scalable systems for a wide range of applications, from communications\cite{smit2019past, xiang2021perspective} and computing\cite{shen2017deep} to sensing\cite{zhang2022large} and metrology\cite{Newman2019Architechture, suh2016microresonator}.
However, some of them demand functionalities beyond what traditional silicon photonics can offer, such as low-loss electro-optic or acoustic-optic modulation, broadband operation down to visible or ultraviolet (UV) wavelengths, and second-order parametric wavelength conversion. 
This has motivated the search for integrated photonics materials with more versatile properties. 

Aluminum nitride (AlN) has emerged as an attractive candidate\cite{li2021aluminium}. 
As a material widely used in the Micro-Electro-Mechanical Systems (MEMS)\cite{kar2011aluminum}, AlN also shows the potential in areas such as optomechanics\cite{xiong2012aluminum} and nonlinear optics\cite{bruch2021pockels}. 
It has a wide bandgap of 6.2 eV\cite{edgar1990low}, allowing operation from deep UV to mid-infrared (MIR)\cite{lu2018aluminum}. Already used as an insulator for microelectronic packaging\cite{liu2023aluminum, kerness1997impurity}, AlN is compatible with complementary metal-oxide semiconductor (CMOS) fabrication processes, making it a promising material for future foundry-level PIC manufacturing. 

Theoretical prediction\cite{takeuchi2002first, farrer2002properties, ranjan2005properties, alsaad2006piezoelectricity, akiyama2009enhancement} of introducing rare-earth dopants for piezoelectricity enhancement has motivated research on doping scandium (Sc) in AlN. 
It has been demonstrated that the piezoeletric coefficient could be improved by more than 5 times in Sc-doped AlN (ScAlN)\cite{akiyama2009enhancement}. 
Further investigation of the effect of Sc doping reveals several favorable properties of ScAlN. 
Sc doping flattens the Gibbs free-energy band structure of the wurtzite AlN, enabling its ferroelectricity\cite{fichtner2019alscn, yang2023domain, wang2023dawn}. 
This allows ferroelectric domain engineering and periodic poling. 
The second-order nonlinearity ($\chi^{(2)}$) has been shown to improve with the Sc content\cite{yoshioka2021strongly}.
The $d_{33}$ component in $36\%$ Sc doped AlN is shown to be one order of magnitude higher than that in the undoped AlN, and also twice of that in lithium niobate (LN). 
Sc doping is thus a promising strategy to improve the nonlinearity of AlN integrated photonic devices. 
However, it also introduces losses and makes the material harder to etch. 
So far, there are very limited studies and successes in achieving low-loss waveguides in ScAlN \cite{zhu2020integrated, yoshioka2021strongly, liu2022ferroelectric}, and few-dB/cm loss is only demonstrated very recently in thin (150 nm) and partially etched (100 nm) ScAlN waveguides, which sacrifices mode confinement in ScAlN\cite{friedman2024measured}. 

In this work, we present a systematic study on the fabrication of low-loss ScAlN PIC. 
We perform a detailed material study on sputter-deposited ScAlN films on 200 mm wafers and develop an optimized fabrication process to produce fully etched waveguides and ring resonators. 
We illustrate the effect of etching and annealing on the resonator quality factor ($Q$). 
By thermally healing the ScAlN thin film, we are able to reduce the absorption loss, achieving an average intrinsic $Q$ over $1.00\times10^5$ (highest at 1.47$\times10^5$, corresponding to 2.4 dB/cm propagation loss). 
Our result paves a viable path towards the development of future low-loss, large-scale, and multifunctional ScAlN PICs.  

\begin{figure*}[!tbh]
\includegraphics[width=1\textwidth]{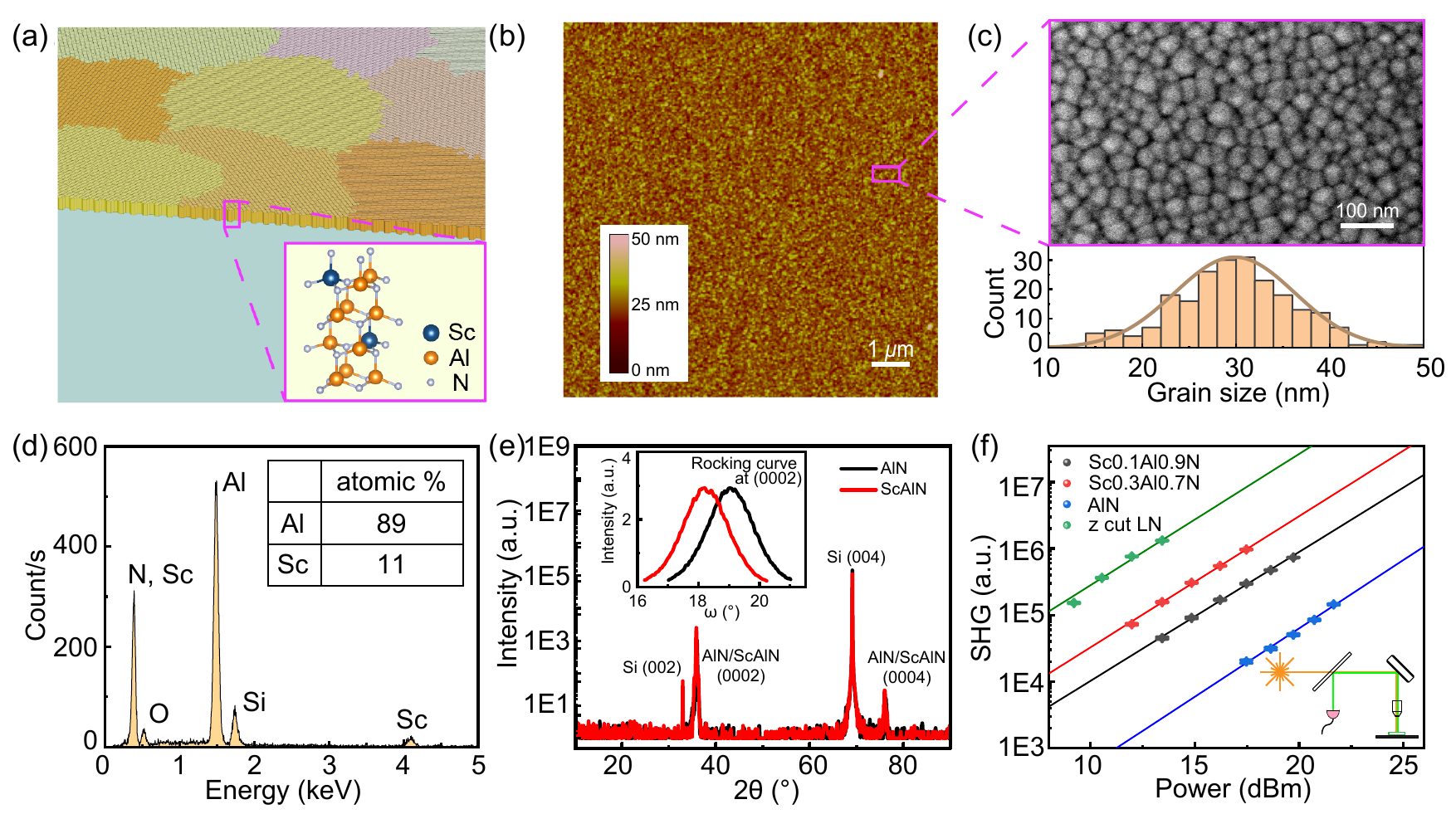}
\caption{\label{fig:1} Material characterization of ScAlN thin films. \textbf{(a)} Illustration of thin-film ScAlN with the $c$-axis perpendicular to the surface. The inset shows the lattice structure of ScAlN. \textbf{(b)} AFM image shows the rough surface after sputtering. The RMS surface roughness is 5.2 nm. \textbf{(c)} SEM image of the thin-film surface shows grain-like structure with a mean grain size of 30 nm. The bottom figure is the grain size distribution. \textbf{(d)} EDX measurement shows the relative elemental composition, revealing a 11$\%$ doping of scandium. The additional silicon and oxygen are from the HSQ residue. \textbf{(e)} The XRD measurement reveals a single phase of sputtered ScAlN and AlN films with the $c$-axis pointing perpendicular to the surface. The slight down-shift of the ScAlN peaks reveals the slight expansion of the lattice due to the larger ionic radius of Sc$^{3+}$ over Al$^{3+}$. The inset is the rocking curve measurements comparing the sputtered AlN and ScAlN. Both have a FWHM of 1.8$^\circ$, indicating the presence of the some amorphous parts. \textbf{(f)} A confocal SHG measurement on thin-film samples demonstrates the nonlinear optical effect in ScAlN. Four samples are 10$\%$ doped ScAlN(Sc$_{0.1}$Al$_{0.9}$N), 30$\%$ doped ScAlN(Sc$_{0.3}$Al$_{0.7}$N), undoped AlN, and z-cut LN. The SHG signal increases with Sc doping concentration. The inset is the schematic of the confocal setup.}
\end{figure*}

\section{Methods and Results}

The ScAlN thin film is prepared by physical vapour deposition (PVD) on 200-mm Si wafers at the Institute of Microelectronics (IME).
A sputtering target of nominal 10$\%$ Sc-doped aluminum is used to deposit 500 nm thick ScAlN films on 4 $\mu$m thermal oxide-on-silicon substrate. 
The ScAlN alloy inherits the lattice structure from either AlN or scandium nitride (ScN), depending on the Sc doping concentration. 
AlN exhibits a stable wurtzite structure (space group P$6_3$mc) with a six-fold in-plane rotation symmetry (point group C$_{6\mathrm{v}}$), and is polar along the $c$-axis. 
ScN has a nonpolar stable ground-state rock-salt phase (space group Fm$\bar{3}$m, point group O$_\mathrm{h}$). 
Theoretical density functional theory (DFT) of ScAlN\cite{zhang2013tunable} predicts a stable wurtzite phase with Sc contents up to 56$\%$, above which the rock salt phase is more favorable. 
The thin-film sample grown from a nominal 10$\%$ scandium doped target is hence expected to be in wurtzite phase. 
Fig. 1(a) shows an illustration of ScAlN thin film on thermal oxide. The honeycomb shape represents the hexagonal shape of the wurtzite structure, when being viewed along the $c$-axis. 
The film is poly-crystalline with individual domains shown by different colors and orientations. 
The inset shows a wurtzite structure with the Al atoms being replaced by the Sc dopants. 
While doping flattens the wurtzite structure towards the layered hexagonal structure\cite{wall2022sputtering}, the current ScAlN film is still expected to exhibit a wurtzite structure due to the relatively low doping concentration.

Fig. 1(b) presents an atomic force microscope (AFM) image of an as-grown film, showing a root-mean-square (RMS) surface roughness of 5.2 nm. 
Congregation of materials forms nanocolumns ranging from <100 nm to 300 nm, evidenced by the bright spots in the AFM image. 
This length scale is comparable to the photon wavelengths of 1550 nm, implying an additional source of scattering loss. 
A magnified image from a scanning electron microscope (SEM) in Fig. 1(c) reveals a granular surface with a mean grain size of 30 nm. 
This provides an avenue for charge trapping, defect creation and dangling bonds, which are all detrimental to film quality. 
The grain size distribution is shown in the lower part of Fig. 1(c). 
An energy-dispersive X-ray spectroscopy (EDX) is performed to identify the elemental composition (see Fig. 1(d)). 
The atomic percentage of the Sc dopants is calculated to be 11$\%$, matching well with the nominal value of the sputtering target. 

To assess the crystallinity and lattice structure information of the ScAlN sample, we compare it with a similarly sputtered undoped AlN thin film using X-ray diffraction (XRD) measurement (see Fig. 1(e)). 
Assignment of planes to peaks is based on the standard diffraction patterns of AlN from crystallographic data base. 
The X-ray wavelength from the synchrotron source is calibrated to be 0.154 nm with the substrate silicon (004) peak at $2\theta = 69.09^\circ$. 
The AlN data shows that it is single-phase with the $c$-axis perpendicular to the film surface. 
The AlN(0002) and AlN(0004) peaks at 35.96$^\circ$ and 76.24$^\circ$ correspond to higher order diffractions (n = 2, 4), and the lattice parameter is calculated to be 0.4989 nm, agreeing well with the literature value of 0.4980 nm\cite{schulz1977crystal}. 
The slight down-shift of the ScAlN(0002) and ScAlN(0004) peaks at 35.84$^\circ$ and 75.96$^\circ$ correspond to a slightly larger lattice parameter of 0.5005 nm, indicating a small expansion of lattice due to the larger ionic radius of Sc$^{3+}$ than that of Al$^{3+}$. 
An additional peak at 32.96$^\circ$ is assigned to the silicon (002) peak. This peak coincides with peak position of the (10-10) peak from AlN/ScAlN with the wurtzite crystal structure. But it is too sharp to be from the thin film. It is most likely from the silicon substrate due to its lattice distortions\cite{zaumseil2015high}.
The rocking curve measurements of the (0002) peaks (Fig. 1(e) inset) show a full-width at half-maximum (FWHM) of 1.8$^{\circ}$ for both doped and non-doped films. 
This value is consistent with other sputter-deposited AlN films in the literature\cite{xiong2012low, iqbal2018reactive}, but larger than those grown using epitaxial growth methods, whose rocking curve FWHM is as small as $0.004^\circ$\cite{zhang2022molecular}. 

Next, we characterize the second-order nonlinear ($\chi^{(2)}$) optical response of ScAlN using a second-harmonic generation (SHG) confocal microscope. 
The inset in Fig. 1(f) shows a simplified schematic of the measurement setup. 
A femtosecond pulsed laser at 1030 nm is focused on the target film through an objective lens. 
The SHG signal at 515 nm is collected through the same objective lens and directed to a silicon avalanche photo-detector (APD) using a dichroic mirror. 
Four different thin-film samples are measured: 10$\%$ doped ScAlN (Sc$_{0.1}$Al$_{0.9}$N), 30$\%$ doped ScAlN (Sc$_{0.3}$Al$_{0.7}$N), undoped AlN, and z-cut LN (shown in Fig. 1(f)). 
The Sc$_{0.1}$Al$_{0.9}$N, Sc$_{0.3}$Al$_{0.7}$N, and AlN films are all 500 nm thick. 
The single-crystal z-cut LN film  (600 nm) is used as a reference to calibrate the $\chi^{(2)}$ values, because the LN is well studied and the ion-sliced single-crystal films have reproducible qualities. The quadratic dependence of the SHG signals on the pump power is illustrated by the straight fitting lines. 
It is clear that the $\chi^{(2)}$ nonlinearity of ScAlN increases with  Sc doping concentration. 
The relevant $\chi^{(2)}$ components for AlN and ScAlN are $d_{33}$ and $d_{31}$. 
In our measurement configuration, since light can only illuminate normally to the film surface, the electric field of the pump light is primarily perpendicular to the crystal $c$-axis. As a result, most SHG signals originate from $d_{31}$. 
Using LN as a reference ($d_{31} = 4.3$ pm/V\cite{zhu2021integrated}), we estimate $d_{31}\simeq$ 0.2, 1.0 and 1.8 pm/V for undoped AlN, 10\% doped ScAlN, and 30\% doped ScAlN, respectively. 
These numbers are on the same order of magnitude as literature values~\cite{kiehne1998optical, yoshioka2021strongly, majkic2017optical, liu2021aln, li2021aluminium}, and the deviations might be caused by differences in film quality and measurement wavelengths. To extract $d_{33}$, angled illumination is required~\cite{yoshioka2021strongly}, which is not accessible in our current setup. During pump power sweep,
we noticed that ScAlN films have higher damage thresholds than AlN, suggesting potentially higher power handling capability with doping.

After the material study, we fabricate waveguides and ring resonators with the 10\% doped ScAlN and assess their losses. 
The ring radius is designed to be 140\,$\mu$m to reduce the bending loss. 
The ring and the bus waveguide widths are 1.5 $\mu$m and 800 nm, respectively. 
A pulley coupling scheme with 5$^\circ$ coupling length is applied to extract the fundamental transverse-electric (TE) mode. 
The coupling gap between the ring and the bus waveguide is varied from 300 nm to 1000 nm to identify the critical coupling condition. 
Grating couplers are used to couple light between the waveguides and optical fibers. 
The devices are patterned using electron beam lithography (EBL) with hydrogen silsesquioxane (HSQ) resist. 
The HSQ patterns are transferred to the ScAlN film using inductively coupled plasma-reactive ion etching (ICP-RIE) with chlorine chemistry~\cite{yan2023development}, and the ScAlN films are fully etched to form ridge waveguides. 
With optimized ICP/RF powers and chamber pressure, the etch selectivity of HSQ to ScAlN is suppressed to <1, and we achieved smooth side wall without re-deposition. Finally, a layer of 1.7 $\mu$m SiO$_2$ cladding is grown on top of the waveguide using plasma-enhanced chemical vapor deposition (PECVD).

\begin{figure*}[!tbh]
\includegraphics[width=0.9\textwidth]{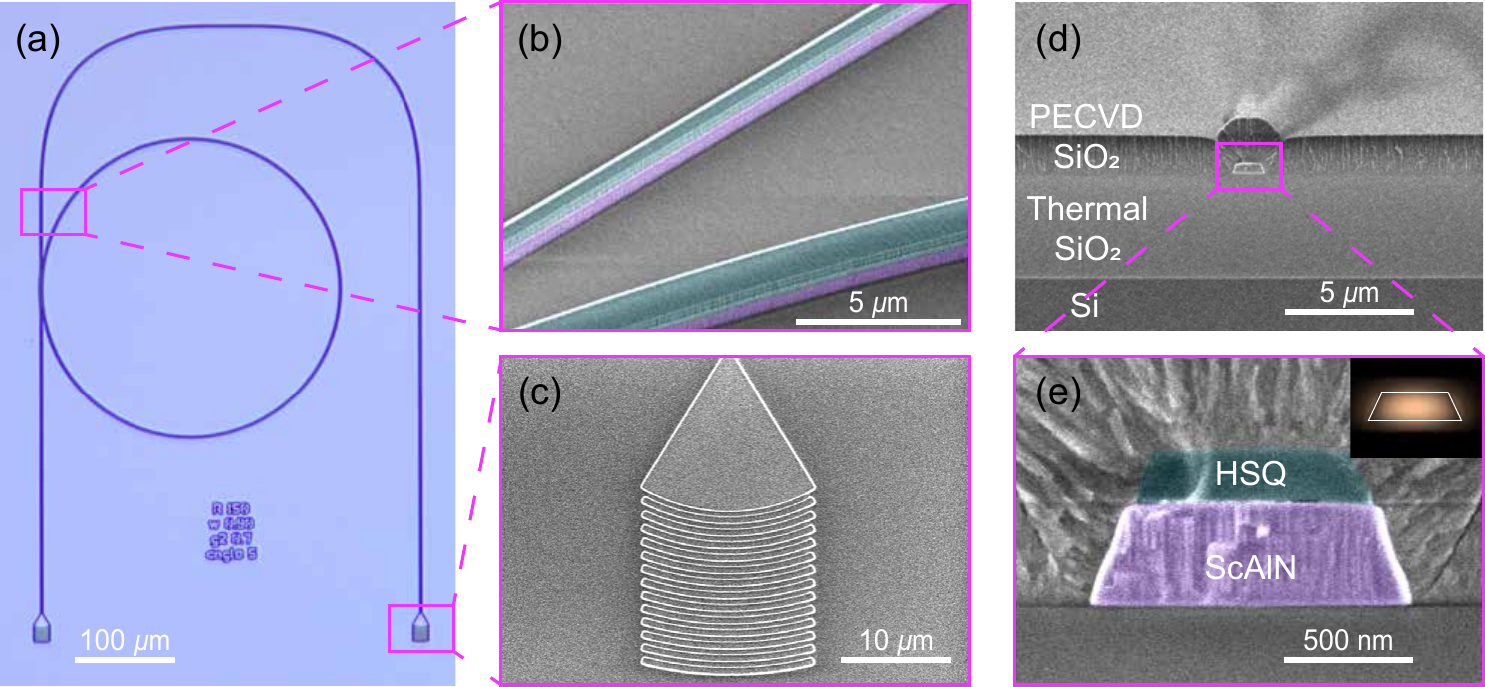}
\caption{\label{fig:2} ScAlN waveguides and ring resonators. \textbf{(a)} Optical microscope image of a typical ring resonator. The ring resonator is pulley-coupled to the bus waveguide. Light in the waveguide is coupled to single-mode fibers using grating couplers. \textbf{(b)} SEM image of ring-bus coupling region. The ScAlN waveguides are false-colored in purple and the HSQ resist is in cyan. \textbf{(c)} SEM image of the grating coupler. \textbf{(d)} SEM image of the waveguide cross-section. A 1.7 $\mu$m-thick PECVD oxide layer is deposited as a cladding. \textbf{(e)} Magnified view of the waveguide cross-section. Inset shows the simulated fundamental TE mode.}
\end{figure*}

Fig. 2(a) shows an optical microscope image of a typical ring resonator. 
Fig. 2(b) is an SEM image of the waveguide, showing a smooth side wall without re-deposition. 
Fig. 2(c) shows the grating coupler, designed to match 8$^\circ$ single-mode fiber array for TE mode coupling at 1550 nm. 
The waveguide cross-section is shown in Fig. 2(d). Fig. 2(e) is a magnified SEM of the waveguide cross-section (ScAlN marked purple, and unremoved HSQ marked cyan), showing a sidewall angle of 70$^\circ$. 
We noticed some vertical strata in the ScAlN waveguide, which may suggest the existence of nanocolumns in the film. 
The inset shows a simulated fundamental TE mode, which is well confined inside the waveguide.

We fabricate 12 devices with various gaps on a single chip to identify the critical-coupling condition.
Fig. 3(a) shows a measured transmission spectrum of the ring resonator from 1500 nm to 1630 nm. 
The extinction ratio is close to 20 dB for most of the fundamental mode resonances, indicating that the resonator is close to critical coupling. 
Several groups of modes are observed. 
By extracting the group indices from the free spectral ranges (FSR), we identify three families of TE modes. 
By fitting the loaded $Q$ and extinction ratio\cite{lu2018aluminum}, we obtain an average intrinsic $Q$ of $1.00 \times 10^5$ for the fundamental TE mode (TE$_0$) from all the resonances within the grating coupler pass band. 
Fig. 3(b) is the intrinsic $Q$ distribution of the TE0 mode over the whole grating coupler pass band. 
The intrinsic $Q$ for the higher-order TE modes (TE1, TE2) are around $0.60 \times 10^5$. 
This is because the higher order modes tend to experience larger interface scattering. 
Fig. 3(c) is a typical resonance near 1550 nm with an intrinsic $Q$ of $1.02 \times 10^5$. 
A maximum intrinsic $Q$ of $1.47 \times 10^5$ is found in a similar device at longer wavelength (Fig. 3(d)). 
This corresponds to a propagation loss of 2.4 dB/cm. 

\begin{figure*}[!htbp]
\includegraphics[width=1\textwidth]{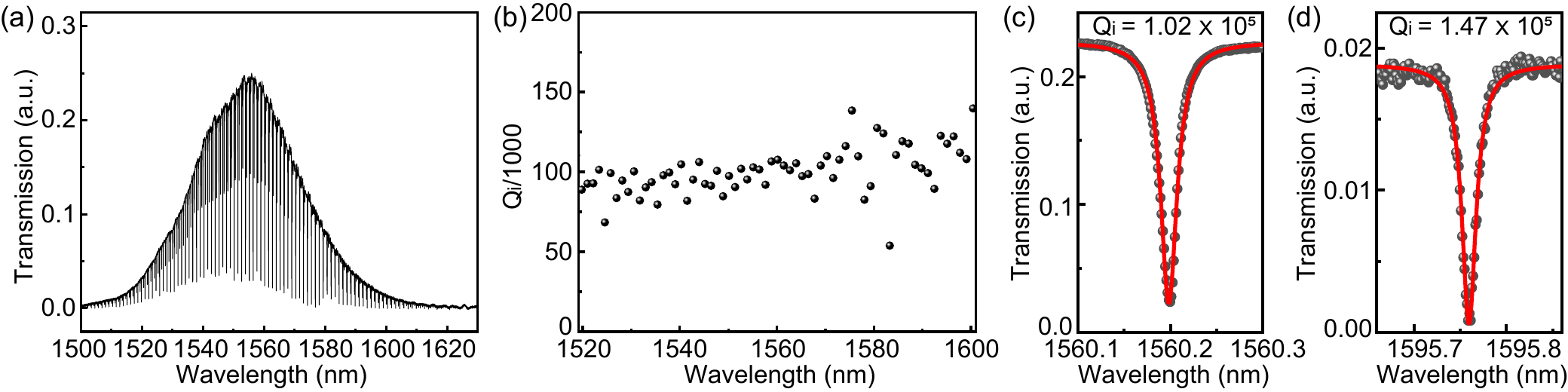}
\caption{\label{fig:3} Transmission measurement and $Q$ factor extraction. \textbf{(a)} Transmission spectrum of a ring resonator with grating couplers. Multiple TE modes are observed. The fundamental TE mode is nearly critically coupled. \textbf{(b)} The average intrinsic $Q$ factors of the fundamental TE mode is around $Q = 1.00\times 10^5$. \textbf{(c)} A typical resonance with $Q = 1.02\times 10^5$. \textbf{(d)} A resonance with an intrinsic $Q = 1.47\times 10^5$ is obtained in a similar device.}
\end{figure*}

\begin{figure*}[htbp!]
\includegraphics[width=0.9\textwidth]{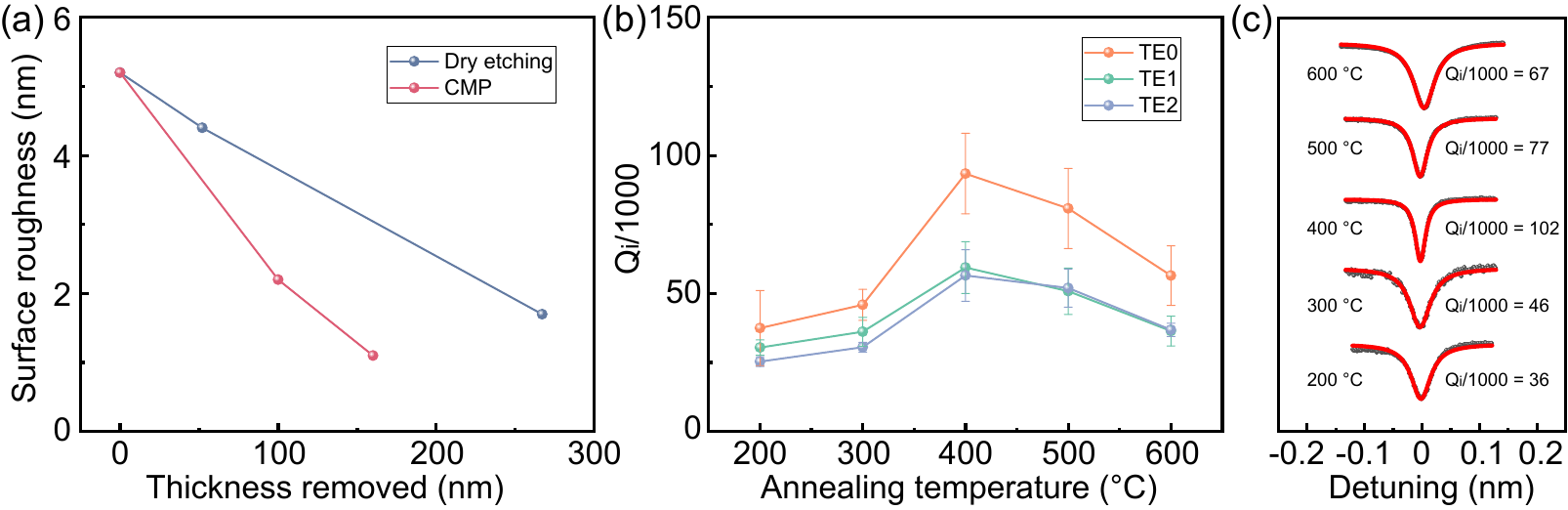}
\caption{\label{fig:4} Loss reduction investigation. \textbf{(a)} Surface roughness against thickness removed for two methods. CMP is more efficient in smoothing the film surface. \textbf{(b)} Intrinsic $Q$ factors as a function of annealing temperatures, revealing an optimal annealing temperature of 400 $^\circ$C. \textbf{(c)} Typical resonances of TE0 mode at different annealing temperatures.}
\end{figure*}

To improve the resonator $Q$ and reduce loss, we have studied several loss-reduction methods during fabrication. 
Two major sources of loss are the scattering loss from the surface and the absorption loss due to material imperfection. 
We investigated the variation of the surface roughness with thickness removed through either dry etching or chemical mechanical polishing (CMP) in Fig. 4(a). 
The CMP method is more efficient in reducing the surface roughness and the final RMS surface roughness is reduced to <1 nm with 160 nm of ScAlN polished away. Compared with the intrinsic $Q$ ($0.5\times10^5$) of the unpolished film, we observed an improvement after polishing.
However, the high-$Q$ devices shown in Fig. 3 are only polished down to 400 nm. 
This is to ensure the mode is mostly confined inside the ScAlN for a maximal confinement factor, which is desired in nonlinear optics applications. 
The material loss is reduced by annealing the device at high temperatures. 
Fig. 4(b) shows the changes of intrinsic $Q$ at different annealing temperatures for 1 h annealing time. The optimal annealing temperature is 400\,$^\circ$C for all TE modes. 
We observed that a longer annealing time was required for more densely written structures, suggesting the possibility of electron-beam damage in the material. 
Any further extension of annealing time does not deteriorate the $Q$. 
Two annealing gases, argon (Ar) and nitrogen (N$_2$), are investigated, but we did not observe a significant difference in the $Q$ factors obtained. 
Fig. 4(c) shows the representative resonances at different annealing temperatures and the corresponding intrinsic $Q$ factors. 
After optimizing the surface roughness and the annealing temperature, the largest potential in improving $Q$ are in reducing material absorption loss and waveguide sidewall scattering loss.

\section{Discussion and Conclusion}
There have been many efforts to vary the deposition conditions, including deposition temperature, power, pressure, gas ratio and substrate\cite{akiyama2009influence, tang2017deposition, fichtner2015stress, zhang2015effects, akiyama2009enhancement, hoglund2010wurtzite} to improve the sputtering film quality. 
Despite the extended efforts, the sputtered ScAlN films generally exhibit a relatively wide rocking curve with a FWHM of >1$^\circ$. 
Hence, other growth techniques such as epitaxial growth with molecular beam epitaxy (MBE)\cite{hardy2017epitaxial, wang2023controlled} and metal organic chemical vapor deposition (MOCVD)\cite{leone2020metal} are being investigated for growing high-quality ScAlN films with better crystallinity. An alternative way to avoid the material absorption is to push the mode outside of the waveguide at the expense of a reduction in mode overlap with the ScAlN waveguide. 
For example, a study on a 30$\%$ doped ScAlN thin film with 150 nm thickness has reported a propagation loss of 2.26 dB/cm from a racetrack resonator measurement, and extracted the corresponding loss from the straight section of the racetrack resonator to be 1.6 dB/cm\cite{friedman2024measured}. In our current work, limited by measurement apparatus and film availability, we were only able to characterize $d_{31}$ at 10\% and 30\% doping concentration. Future upgrade of the measurement apparatus to include variable angle, multi-wavelength, and polarization resolved measurement is necessary to assess the full $\chi^{(2)}$ tensor at different wavelengths, especially $d_{33}$, which is expected to be the largest element~\cite{yoshioka2021strongly}. In addition, a larger range of Sc doping concentrations needs to be tested to get an empirical relation between nonlinearity and doping concentration. We expect more relevant works to come out in the coming years to help populate this database.

In summary, we have demonstrated the fabrication of low-loss waveguides and ring resonators on PVD-grown thin-film ScAlN. 
The highest measured intrinsic $Q$ is $1.47\times10^5$, corresponding to a propagation loss of 2.4 dB/cm. 
A dry etching recipe based on chlorine chemistry is developed to achieve a fully etched structure with a smooth, redeposition-free sidewall. 
We identified an optimal annealing temperature of 400\,$^\circ$C for reducing the waveguide loss. 
Our results serve as an important step towards future large-scale photonic integrated circuits on thin-film ScAlN. 
Currently, the dominant loss is still from material imperfections, which may be improved through further optimizations in deposition processes to achieve better crystallinity and homogeneity. 
Future developments of functional devices beyond linear components, such as electro-optic modulators, acousto-optic modulators, and various nonlinear optical devices, are the immediate next steps to unlock the full potential of this material platform. 

\textit{Note}: during the paper revision, we noticed another work posted online reporting similar intrinsic $Q$ of $1.4\times10^5$ in ScAlN~\cite{yang2024unveiling}. 

\section{Acknowledgement}
This work is supported by A*STAR (C230917005 for device fabrication and measurement, C220415015 for ScAlN thin film deposition, and M23M5a0069 for ScAlN characterization and results analysis) and National Research Foundation (NRF2022-QEP2-01-P07, NRF-NRFF15-2023-0005, NRF-CRP28-2022-0002). The authors would like to thank Jie Deng, Seng Kai Wong, Sherry Yap, Aihong Huang and Qingxin Zhang for assistance in device fabrication; Ming Lu and Jiapeng Sun for assistance in manuscript preparation. The fabrication of the devices was done at the A$^\ast$STAR Cleanroom.

\section{Author Declarations}
The authors have no conflicts to disclose.\\

\noindent\textbf{REFERENCES}

\bibliography{References}

\end{document}